\documentclass[aps,prl,twocolumn,showpacs,superscriptaddress,groupedaddress]{revtex4}  % for review and submission
\usepackage{graphicx}  % needed for figures
\usepackage{dcolumn}   % needed for some tables
\usepackage{bm}        % for math
\usepackage{amssymb}   % for math
\usepackage{footnote}
\usepackage{multirow}
\usepackage[hyperindex,breaklinks]{hyperref} 
\usepackage{atlasphysics}

\usepackage{preprintcover}  % load package
 \PreprintCoverPaperTitle{Search for $tb$ resonances in proton-proton collisions \\ at $\sqrt{s}$ = 7~\TeV~with the ATLAS detector}  % paper title
 \PreprintIdNumber{CERN-PH-EP-2012-087}  % CERN preprint number
 \PreprintCoverAbstract{This Letter presents a search for $tb$ resonances in 1.04~$\mathrm{fb}^{-1}$ of 
 LHC proton-proton collision data collected by the ATLAS detector at a center-of-mass energy of 7~\TeV. 
 Events with a lepton, missing transverse momentum, and two jets are selected and the invariant mass 
 of the corresponding final state is reconstructed. 
 The search exploits the shape of the $tb$ invariant mass distribution compared to the expected Standard Model backgrounds.
 The model of a right-handed $W'_R$ with Standard Model-like couplings is chosen as the benchmark model for this search.
 No statistically significant excess of events is observed in data, and upper limits on the cross section times the 
 branching ratio of $W'_R$ resonances at 95$\%$ CL lie in the range 6.1--1.0 pb 
 for $W'_R$ masses ranging from 0.5 to 2.0~\TeV. These limits are translated into a lower bound on the allowed 
 right-handed $W'_R$ mass, giving $m_{W'_R} >$  1.13~\TeV~at 95$\%$ CL.
 }  % paper abstract
 \PreprintJournalName{Physical Review Letters}  % journal name
\clearpage
\begin{document}

\title{Search for $tb$ resonances in proton-proton collisions at $\sqrt{s}$ = 7~\TeV \\ with the ATLAS detector}

\author{The ATLAS Collaboration}

\begin{abstract}
This Letter presents a search for $tb$ resonances in 1.04~$\mathrm{fb}^{-1}$ of 
LHC proton-proton collision data collected by the ATLAS detector at a center-of-mass energy of 7~\TeV. 
Events with a lepton, missing transverse momentum, and two jets are selected and the invariant mass 
of the corresponding final state is reconstructed. 
The search exploits the shape of the $tb$ invariant mass distribution compared to the expected Standard Model backgrounds.
The model of a right-handed $W'_R$ with Standard Model-like couplings is chosen as the benchmark model for this search.
No statistically significant excess of events is observed in data, and upper limits on the cross section times the 
branching ratio of $W'_R$ resonances at 95$\%$ CL lie in the range 6.1--1.0 pb 
for $W'_R$ masses ranging from 0.5 to 2.0~\TeV. These limits are translated into a lower bound on the allowed 
right-handed $W'_R$ mass, giving $m_{W'_R} >$  1.13~\TeV~at 95$\%$ CL.

%%%%%%%%%%%%%%%%%%
% SUMMARY
%%%%%%%%%%%%%%%%%%

\end{abstract}

\maketitle
%\linenumbers

%%%%%%%%%%%%%%%%%%%%%%%%%%%%%%%%%%%%%%%%%%%%%%%%%%%%%%%%%%%%%%%%
%
%                    INTRODUCTION
%
%%%%%%%%%%%%%%%%%%%%%%%%%%%%%%%%%%%%%%%%%%%%%%%%%%%%%%%%%%%%%%%%

This Letter presents a search for $tb$ ($t\bar{b}$ or $\bar{t}b$) resonances using data collected in 2011 by the ATLAS detector~\cite{ATLAS_det} at the Large Hadron Collider 
(LHC), corresponding to an integrated luminosity of 1.04~$\pm$~0.04 $\mathrm{fb}^{-1}$~\cite{Lumi, Lumi2}
from $pp$ collisions at a center-of-mass energy of 7~\TeV. 
These resonances include new heavy gauge bosons such as the $W'$ boson. 
The $W'$ boson is a charged heavy gauge boson that is predicted in many extensions of the Standard Model (SM) such as 
universal extra dimensions~\cite{extradimensions} and little Higgs models~\cite{little-Higgs}. If the $W'$ boson is assumed to have similar coupling 
strengths to those of the SM $W$ boson, searches in the $W'  \rightarrow \ell\nu$ decay channel, 
where $\ell$ is a charged lepton, are the most sensitive. 
However, the $W' \rightarrow tb$ channel is competitive if $W' \rightarrow \ell\nu$ decay is suppressed. For example, 
for a right-handed $W'_R$ this can happen if the right-handed neutrino, $\nu_R$, is heavy enough to prevent 
$W'_R \rightarrow \ell \nu_R$ decay~\cite{extended-gauge-model}.
The model of a right-handed $W'_R$ with SM-like couplings is chosen as the benchmark model for the analysis presented in this Letter.
The  $W'_R \rightarrow tb$  decay channel has been searched for at the Tevatron~\cite{Wprime_D0,Wprime_CDF}. 
The best previous limit on a $W'_R$ with Standard Model-like couplings of the $W'$ to quarks was set by the D0 experiment 
and excludes a $W'_R$ mass below 890~\GeV~at 95$\%$ confidence level.
%%%%%%%%%%%%%%%%%%%%%%%%%%%%%%%%%%%%%%%%%%%
%
%   ATLAS DETECTOR
%
%%%%%%%%%%%%%%%%%%%%%%%%%%%%%%%%%%%%%%%%%%%

The innermost part of the ATLAS detector~\cite{coordinates}, a tracking system in a 2~T axial magnetic field, measures the momentum of
the charged particles produced in the collisions. Outside of the solenoid are the 
calorimeter subsystems, which measure the electron, photon, and hadronic particle energies,
and the muon spectrometer which is used to identify and measure the momentum of muons in a toroidal magnetic field.
A three-level trigger system~\cite{ATLAS_trigger} reduces the event rate and selects the events for analysis. 

The $tb$ resonances are searched for in the $tb \rightarrow \ell \nu bb$ decay 
channel, where the lepton, $\ell$, is either an electron or a muon. 
%%%%%%%%%%%%%%%%%%
% Simulation
%%%%%%%%%%%%%%%%%%
$W'_R$ signal events are simulated to leading order (LO) with the {\sc PYTHIA} v6.421~\cite{pythia} Monte Carlo (MC) generator, 
using the MRST2007 LO* parton distribution functions (PDFs)~\cite{MRST2007}. 
Seven signal samples are simulated, with different $W'_R$ mass assumptions, ranging from 500~\GeV~to 2.0~\TeV, 
as reported in Table~\ref{tab:signal_xs}.
The respective signal cross section times the branching ratio values are computed 
at next-to-leading-order (NLO)~\cite{ZackSullivan}, using CTEQ6.6 PDFs~\cite{cteq6l}. 
 
 \begin{table}[h]
 \begin{ruledtabular}
     \begin{tabular}{r@{\extracolsep{\fill}}r@{\extracolsep{0pt}$\:\pm\:$}l@{}%
         @{\extracolsep{\fill}}r@{\extracolsep{0pt}$\,\pm\,$}l@{}}
       $m_{W'_R}$ [\GeV]    &\multicolumn{2}{c}{${\cal {B}}(W'_R \rightarrow  tb)$}  & \multicolumn{2}{c}{$\sigma \times {\cal {B}}$} [pb] \\ \hline \hline
        \phantom{0}500 &  0.298& 0.002     & 54.6  &2.1  \\	
        \phantom{0}750 &  0.319& 0.001     & 10.9  &0.6 \\	
       1000            &  0.326& 0.001     &  2.92 &0.18 \\	
       1250            &  0.328& $<$ 0.001 &  0.91 &0.07 \\	
       1500            &  0.330& $<$ 0.001 &  0.31 &0.03 \\	
       1750            &  0.331& $<$ 0.001 &  0.11 &0.01 \\	
       2000            &  0.332& $<$ 0.001 &  0.044&0.005 \\%0.04 $\pm$ 0.01	
     \end{tabular}       
 \end{ruledtabular}
   \caption{\label{tab:signal_xs} NLO branching ratios, ${\cal{B}}(W'_R \rightarrow  tb)$, and $W'_R$ production cross section 
     times the branching ratio value, $\sigma(pp \rightarrow W'_R) \times {\cal{B}}(W'_R \rightarrow  tb)$, in $pp$ collisions at 7~\TeV~center-of-mass energy~\cite{ZackSullivan}.  The uncertainties on the branching ratios are due to the top quark mass 
     uncertainty. The uncertainties on the cross sections include statistical, 
     $\alpha_s$, NLO renormalization and factorization scales, and PDF uncertainties. }
 \end{table}
 
Data-driven methods and MC simulated samples are used to estimate and model backgrounds.
The $t\bar{t}$ process is simulated with the {\sc MC@NLO} 
v3.41~\cite{SAMPLES-MCNLO1,SAMPLES-MCNLO2} MC generator,
assuming a top quark mass of 172.5~\GeV, and using the CTEQ6.6 PDFs. The parton shower is added using the {\sc HERWIG}~\cite{SAMPLES-HERWIG} and 
{\sc JIMMY}~\cite{JButterworth:1996zw} MC generators. The $t\bar{t}$ cross section is obtained from the approximate 
NNLO prediction calculated with the \textsc{HATHOR} program~\cite{Aliev:2010} using the MSTW2008 NNLO PDF sets~\cite{Martin:2009iq}. 
The single top quark processes are simulated using 
the {\sc AcerMC} v3.7~\cite{SAMPLES-ACER} MC generator and hadronization is performed with the \textsc{PYTHIA} MC generator; the cross section is calculated to approximate 
NNLO~\cite{Kidonakis:2010ux,Kidonakis:2010tc,Kidonakis:2011wy} using the CTEQ6.6 PDFs. Diboson processes are 
simulated using the {\sc HERWIG} v6.5 MC generator and their cross sections are obtained 
at NLO using the MCFM~\cite{NLO-MCFM} program with the MSTW2008 PDFs. 
The MC samples simulated with the {\sc AcerMC} and {\sc HERWIG} MC generators use the MRST2007 LO* PDFs. Vector boson 
production in association with jets ($W+$ light jets, $Wb\bar{b}$, $Wc\bar{c}$, $Wc$ and $Z+$jets with up to five additional partons) 
is simulated using the {\sc ALPGEN} v2.13~\cite{SAMPLES-ALPGEN} MC generator, coupled with the CTEQ6L1 PDFs~\cite{cteq6l} and hadronization is performed with the {\sc HERWIG} and {\sc JIMMY} MC generators.  
In these samples, additional jets can be created from the parton shower.
In order to avoid double counting between the inclusive $W+n$ parton samples 
and the parton shower, overlaps are removed following the MLM 
matching prescription~\cite{SAMPLES-ALPGEN}.
A cross section correction factor  is applied to the LO $W$/$Z$+jet cross sections computed by comparing the LO and NLO predictions from the \textsc{FEWZ}~\cite{Anastasiou:2003ds} program. The $Wc$ cross section correction factor is obtained using the \textsc{MCFM}~\cite{Campbell} 
program with the CTEQ6.6 PDFs.
All samples are passed through 
the full simulation of the ATLAS detector~\cite{ATL-2010-005} 
based on {\sc GEANT4}~\cite{SAMPLES-G4} and are then reconstructed using the same procedure as collision data. 
The simulated samples include the effect of multiple $pp$ collisions per 
bunch crossing (pile-up) which on average is six events per bunch crossing.
In order to ensure a good description of the energy scale and resolution, the trigger, the reconstruction and identification efficiency, 
corrections based on comparisons  between data and MC events are applied to the simulated signal and background samples. 
The corresponding scale factors are obtained as a function of the object kinematics, resulting 
in final corrections of the order of a few percent.

%%%%%%%%%%%%%%%%%%
% Final State
%%%%%%%%%%%%%%%%%%

Candidate events are identified using single 
high transverse momentum electron and muon triggers  and stringent 
detector and data quality requirements.
For each candidate, two jets, one isolated charged lepton, and missing transverse 
momentum, $E_\mathrm{T}^{\mathrm{miss}}$, are required. 
%%%%%%%%%%%%%%%%%%%%%%%%%%%%%%%%%%%%
%         Event Selection
%%%%%%%%%%%%%%%%%%%%%%%%%%%%%%%%%%%%
The definition of the objects and details of a similar event selection including lepton isolation requirements are given in Ref.~\cite{objects}.
The reconstructed charged lepton is required to have a transverse momentum $p_\mathrm{T}>$ 25~\GeV~to ensure a constant 
trigger efficiency, $|\eta|<$ 2.5 for a muon~\cite{muon_trig, muon_reso, muon_reco} and $|\eta|<$ 2.47 for an electron~\cite{ele_perf} 
(the calorimeter transition region 1.37 $<|\eta|<$ 1.52 is excluded), and to lie within 
$\Delta R=\sqrt{(\Delta\eta)^2+(\Delta\phi)^2}<$ 0.15 of the corresponding triggered lepton.
Jets are reconstructed from energy clusters in the calorimeters with the anti-$k_\mathrm{t}$ 
algorithm~\cite{Salam:2007xv} with a radius parameter $R=0.4$ and calibrated to the hadronic energy scale~\cite{jetcali}. 
Exactly two jets with $p_\mathrm{T}>$ 25~\GeV~and $|\eta|<$ 2.5 are required in the event, and at least one of them must be 
tagged as a $b$-jet. 
The $b$-tagging algorithm uses measurements of the impact parameters of tracks and the properties 
of reconstructed vertices; these are combined in a neural network to extract a tagging decision for each jet~\cite{ATLAS-CONF-2011-102}.
Based on a $t\bar{t}$ MC sample, the working point is chosen at a $b$-tagging 
efficiency of 57$\%$, leading to a light-quark tagging probability of 0.2$\%$ derived from the same sample. 
To account for the differences between observed and simulated jet $p_{\mathrm{T}}$ and $\eta$ distributions, 
the $b$-tagging efficiency and the corresponding scaling factors to be applied to MC are derived from data~\cite{ATLAS-CONF-2011-089}. 
Events before applying any $b$-tagging are referred to  as {\it pre-tagged} events.
Events where one or both jets are $b$-tagged are referred to as {\it single-} or {\it double-tagged} events respectively. 

The $E_\mathrm{T}^{\mathrm{miss}}$ is calculated using calorimeter energy clusters~\cite{clusters} calibrated according 
to the reconstructed physics object to which they are associated~\cite{ATLAS-MET}; 
events are required to satisfy $E_\mathrm{T}^{\mathrm{miss}} > $ 25~\GeV.
The background contribution from multiple hadron jets (multijet background) is 
reduced by imposing  a requirement on the sum of the $W$ boson transverse mass 
$m_{\mathrm{T}}(W)$~\cite{mtW} and $E_\mathrm{T}^{\mathrm{miss}}$: $m_{\mathrm{T}}(W) + E_\mathrm{T}^{\mathrm{miss}}>~60~\mathrm{\GeV}$~\cite{reduceQCD}. 
After applying all selection criteria, the acceptance times efficiency for $W'_R$ signal events with 
$m_{W'_R}$ = 1.0~\TeV~is 1.38$\%$  for single-tagged events and 0.49$\%$ for double-tagged events.

%%%%%%%%%%%%%%%%%%
% Bkg composition
%%%%%%%%%%%%%%%%%%
One of the most important backgrounds for the $tb$ resonance search comes from $W$ production in association 
with either heavy-flavor jets, or light-flavor jets misidentified as $b$-jets.   
Multijet production is another source of background, when 
either a hadronic jet is misidentified as
a lepton, or when a real high-$p_\mathrm{T}$ lepton from semileptonic decay of a 
heavy hadron within a jet fulfills the selection requirements.
Another important background comes from $t\bar{t}$ pair
production in the case that one $W$ boson decays leptonically and the decay products of the other 
$W$ boson are lost due to the detector acceptance.
Other smaller backgrounds come from single top production, diboson production and 
$Z+$jet events. 
%%%%%%%%%%%%%%%%%%
% Bkg estimation
%%%%%%%%%%%%%%%%%%t

Kinematic variable distributions for the $W$+jet background are taken from MC samples,
while the overall normalization and flavor composition are derived from data; 
this is done after rejecting signal-like events with the $tb$ invariant mass, 
$m_{tb}$ which is described later, satisfying $m_{tb}>$ 500~\GeV. In each jet multiplicity bin, the number of $W$+jet events 
in the data is assumed to be the difference between the number of observed data events and 
the number of events estimated for SM non-$W$+jet processes including the multijet process
estimated from a data-driven method. 
The overall $W$+jet normalization factor is the ratio of the number of $W$+jet
events in the data to the number of $W$+jet events in simulation.
The flavor composition of the $W$+jet background is estimated by comparing the MC prediction 
to data while its dependence on jet and $b$-tagging multiplicity is modeled using MC. The fractions 
of $Wb\bar{b}$, $Wc\bar{c}$, $Wc$ and $W$+light jet components of the total $W$+jet 
MC are scaled such that the background sum equals the observed data in three separate samples: 
a single-tagged one-jet sample and the pre-tagged and single-tagged two-jet samples. The same 
scale factor is used for  $Wb\bar{b}$ and $Wc\bar{c}$.

The multijet background normalization and the shape of each distribution are obtained from data.
The shape of each multijet background distribution is taken from a data sample which requires a jet instead of an isolated lepton. 
This jet is required to have a detector signature similar to an electron: it must have $p_\mathrm{T} > $ 25~\GeV~and between 80\% and 95\% of its 
energy deposited in the electromagnetic section of the calorimeter.
The jet must also be associated with at least four tracks.
The normalization is estimated using a binned likelihood fit to the $E_\mathrm{T}^{\mathrm{miss}}$~distribution in data in 
which the normalization of the $W$+jet and the multijet components is allowed to vary. %~\cite{ATLAS-CONF-2011-027}.
The fit is performed separately in the pre-tagged, single- and double-tagged samples,
after applying all selection criteria except the $E_\mathrm{T}^{\mathrm{miss}}$~cut. 
The uncertainty on the multijet rate is 50\% for pre-tagged and single-tagged events, while it amounts to 100\% for double-tagged events.
The uncertainty is estimated by using the $m_{\mathrm{T}}(W)$ distribution 
instead of the $E_\mathrm{T}^{\mathrm{miss}}$ distribution in the binned 
likelihood fit, and by using multijet background models built from data samples with low and high numbers of $pp$ collisions per event.

The $t\bar{t}$, single top, $Z$+jet and diboson events are normalized to the theoretical cross sections and the shape of each distribution is taken from the MC simulation.

Based on the theoretical predictions shown in Table~\ref{tab:signal_xs}, the numbers of single- and 
double-tagged $W'_R$ signal events expected in 1.04~$\mathrm{fb}^{-1}$  are listed in Table~\ref{tab:sg_pretag}, as a function of $m_{W'_R}$.
Table~\ref{tab:evtyield_pretag}  lists the expected background yields.
\begin{table}[h]
\begin{ruledtabular}
\begin{tabular}
      {r@{\extracolsep{\fill}}r@{\extracolsep{0pt}$\:\pm\:$}l@{}%
        @{\extracolsep{\fill}}r@{\extracolsep{0pt}$\,\pm\,$}l@{}}
    $m_{W'_R}$ [\GeV]    & \multicolumn{2}{c}{Single-tagged} & \multicolumn{2}{c}{Double-tagged}  \\ \hline \hline

     \phantom{0}500      & 973   & 37     &  455  & 17 \\    
     \phantom{0}750      & 174   & 9      &  77   & 4 \\      
    1000     &  42   & 3      &  15   & 1 \\     
    1250     &  11   & 1      &  3.9  & 0.3 \\      
    1500     &  3.2  & 0.3    &  1.0  & 0.1 \\      
    1750     &  1.0  & 0.1    &  0.26 & 0.03 \\
    2000     &  0.36 & 0.04   &  0.09 & 0.01 \\
  \end{tabular}
\end{ruledtabular}
  \caption{Predicted signal event yields derived using the theoretical cross section times the branching ratio values for $W'_R \rightarrow  tb$, 
  for single- and double-tagged two-jet events in 1.04~$\mathrm{fb}^{-1}$ of data.
   The uncertainties correspond to the NLO cross section uncertainties~\cite{ZackSullivan}.
\label{tab:sg_pretag}}
\end{table}
 \begin{table}[!h!btp]
 \begin{ruledtabular}
 \begin{tabular}
       {l@{\extracolsep{\fill}}r@{\extracolsep{0pt}$\:\pm\:$}l@{}%
         @{\extracolsep{\fill}}r@{\extracolsep{0pt}$\,\pm\,$}l@{}}
 Samples & \multicolumn{2}{c}{Single-tagged} & \multicolumn{2}{c}{Double-tagged}
                                         
 \vspace{0.03in} \\
 \hline
       $W$+ jets	               &  5970 & 1000               &   290 & 180  \\ 
       Multijets		       &  1120 & 560                &    47 & 47   \\ 
       $t\bar{t}$		       &  1560 & 130                &    360 & 30  \\ 
       Single top                      &  1240 & 90                 &    120 & 10  \\ 
       Diboson, Z+jets		       &   320 & 120                &    14 & 2    \\ 
       \hline
       Total prediction		        & 10200 & 1200               &   830 & 190  \\ 
       Data                      & \multicolumn{2}{c}{10428}
       & \multicolumn{2}{c}{844~~~~~~~~~}
 \end{tabular}
 \end{ruledtabular}
   \caption{Predicted background event yields compared to the total observed event yields for single- and double-tagged two-jet events in 
     1.04~$\mathrm{fb}^{-1}$ of data.  All $W$+jet samples are scaled by the
     factors determined from data, with the uncertainties also derived from data. 
     The multijet estimation is from the fitting method with a 50$\%$ (100$\%$) uncertainty 
     for single(double)-tagged events. 
     All the other predictions are derived using the theoretical cross sections and 
     uncertainties.
     \label{tab:evtyield_pretag}}
 \end{table}
 
%%%%%%%%%%%%%%%%%%
% ANALYSIS
%%%%%%%%%%%%%%%%%%
The $tb$ invariant mass is used as the observable to discriminate signal from background. 
The neutrino momentum  in the decay $tb\rightarrow \ell \nu bb$ is computed assuming the transverse
component to be equal to $E_\mathrm{T}^{\mathrm{miss}}$, and extracting the longitudinal component ($p_z$) by constraining the 
 $\ell$-$\nu$ invariant mass to $m_W$ = 80.42~\GeV. This gives a quadratic equation in $p_z$ and the 
solution with the smaller $|p_z|$ is used. If the solution is complex, only the real part is taken and the imaginary part is neglected.

\begin{figure}[h]
 \centering
 \includegraphics[width=0.5\textwidth]{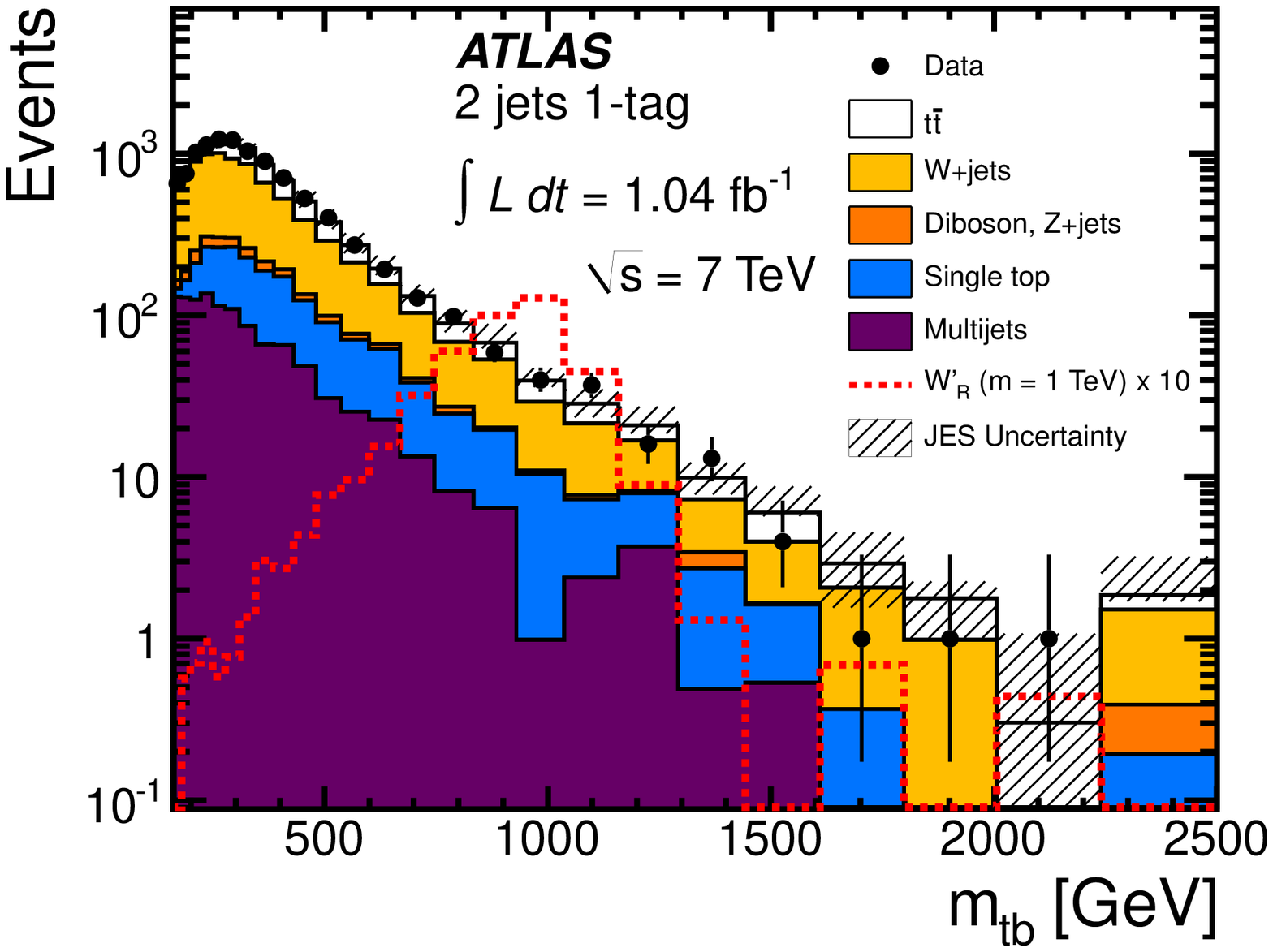}
 \includegraphics[width=0.5\textwidth]{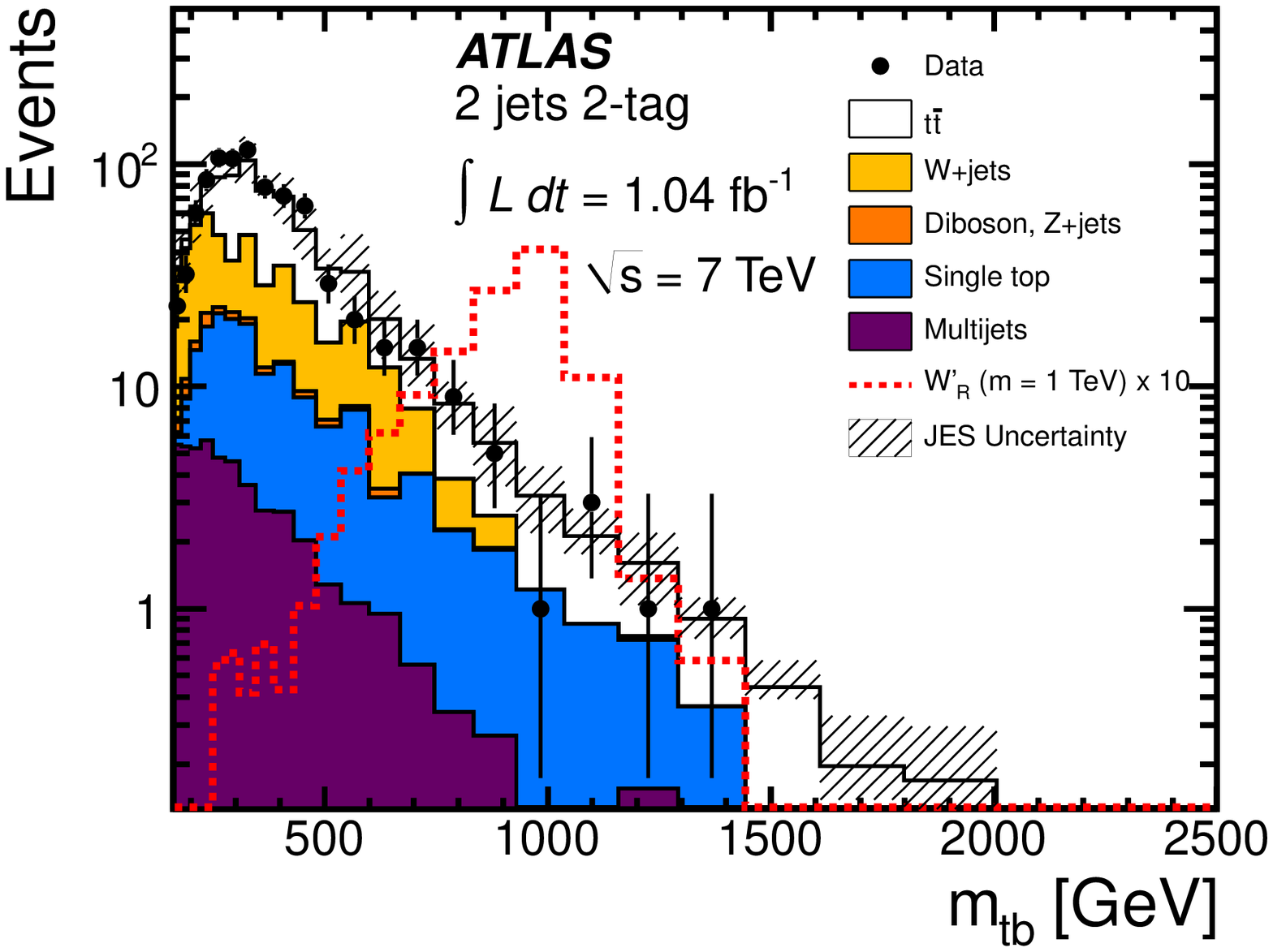}
 \caption{The distribution of $m_{tb}$ for single-tagged (top) and double-tagged (bottom) two-jet events in data compared to Standard Model expectations. The expected $W'_R$ signal, normalized to the theoretical cross section times the ${\cal{B}}(W'_R \rightarrow  tb)$ values from Table~\ref{tab:signal_xs}, has been scaled by a factor of 10. The effect of the jet energy scale (JES) uncertainty on the predicted background is shown, as are the data 
statistical uncertainties. The bin width is constant in log($m_{tb}$). The highest bin in each plot includes overflows.}
 \label{fig:invmass_2tag}
 \end{figure}

Figure~\ref{fig:invmass_2tag} shows the data and expected background distributions of $m_{tb}$ for single- and double-tagged two-jet events.
The data event with the highest $m_{tb}$ value corresponds to a single-tagged event with $m_{tb} \simeq $ 2.0~\TeV.
The {\sc Bumphunter} tool~\cite{bumphunter}  is used  to search for a local excess in the data due to the production of a $tb$ resonance.
This tool is used to test the consistency of the data with the SM background only hypothesis, comparing 
the data to the SM prediction over the spectrum of the $tb$ invariant mass,
scanning over sliding mass windows from 0.5 to 2.0~\TeV. 
The width of the mass windows is chosen to be constant in log($m_{tb}$) as shown in Fig.~\ref{fig:invmass_2tag} to deal with low 
background MC statistics in the higher mass bins.
This comparison has been performed for single- and double-tagged events separately. The region 
 with the highest data-background difference is 1024--1129 (764--842)~\GeV~for 
single(double)-tagged events. The probability of observing the SM background fluctuating up to or above 
the number of observed data events in these regions is 0.66 for single-tagged events and 0.72 for 
double-tagged events. %, which translates into a significance of 0.60 (0.58) $\sigma$. 
These values, which are based on the statistical error only, indicate that there is no significant 
evidence for $tb$ resonances in the observed data.

%%%%%%%%%%%%%%%%%%%%%%%%%%%%%%
%%% Systematics
%%%%%%%%%%%%%%%%%%%%%%%%%%%%%%
Systematic uncertainties  from various sources affecting the background and the signal acceptance (rate uncertainty), 
as well as shape changes in the invariant mass distribution (shape uncertainty) are considered. 

The jet energy scale and the uncertainty on the $b$-tagging scale factors are the dominant systematic uncertainties 
for the signal. The background normalization yields are the dominant systematic uncertainty for the background contribution.
The jet energy scale uncertainty is evaluated by scaling 1$\sigma$ up or down the energy of each jet.
The $b$-tagging scale factors are $p_\mathrm{T}$-dependent and have an uncertainty between 8$\%$ and 20$\%$.
The multijet background uncertainty  has already been described.
The uncertainty on the normalization of the $W$+jets background and its flavor composition
include both systematic contributions and  a statistical contribution from the limited size of the sample.
The $W$+jet flavor uncertainties are treated as fully correlated between $Wb\bar{b}$ and
$Wc\bar{c}$ and uncorrelated otherwise.
Theoretical cross section uncertainties for the top ($t\bar{t}$ and single top), diboson 
and  $Z+$jet backgrounds of 10\%, 5\% and 60\% are assigned, respectively. 
The $Z+$jet theoretical cross section uncertainty is estimated  based on the variation of 
\textsc{ALPGEN} parameters, and a relative uncertainty of 50\% on the heavy-quark 
contributions, but it has a very small impact on the result due to the small contribution of $Z$+jet events.
Systematic uncertainties due to the residual differences between data and 
MC simulation for the reconstruction and energy calibration of 
jets, electrons and muons are estimated to have
a small impact on the result.
The uncertainty on the integrated luminosity is 3.7\%~\cite{Lumi2}.
The uncertainty on the background modeling in the $m_{tb}$ distribution is evaluated using pre-tagged data and found to be negligible.

An uncertainty due to the MC event generator is estimated by comparing {\sc MC@NLO} and {\sc POWHEG}~\cite{SAMPLES-POWHEG, SAMPLES-POWHEG-1} 
for $t\bar{t}$ and {\sc AcerMC} and {\sc MC@NLO} for single top events. 
The uncertainty in parton shower modeling is estimated by comparing two {\sc POWHEG} $t\bar{t}$ samples for which 
the hadronization is performed by {\sc PYTHIA} or {\sc HERWIG}.  
Uncertainties from modeling the amount of initial and final-state QCD radiation are also taken into account.
The uncertainty due to the specific choice of PDFs in the simulated events is determined by re-weighting the MC events 
using the NNPDF20, MSTW2008 and CTEQ6.6~\cite{Martin:2009iq} eigenvector PDF sets. 
Finally, an uncertainty to account for the limited MC sample sizes is also included.

\begin{figure}[h]
 \centering
 \includegraphics[width=0.5\textwidth]{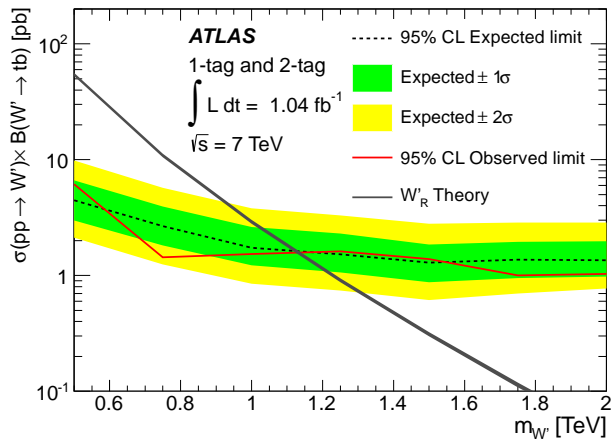}
  \caption{95$\%$ CL limit on the cross section, $\sigma(pp \rightarrow W'_R)$, times branching ratio for $W'_R \rightarrow tb$  as a 
function of the $W'$ boson mass. The theory curve is also shown.
 \label{fig:limitsys}}
 \end{figure}

%%%%%%%%%%%%%%%%%%%%%%%%%%%%
% Results
%%%%%%%%%%%%%%%%%%%%%%%%%%%%
No significant data excess is identified for any value of $m_{tb}$,
and an upper limit on the $W'_R \rightarrow tb$ production cross section ($\sigma$) times 
the ${\cal{B}}(W'_R \rightarrow  tb)$ at 95\% credibility-level (CL) is determined using 
a Bayesian approach assuming flat priors~\cite{Bayesian}.
The likelihood function used is the product of the Poisson probabilities over all mass 
bins~\cite{Zwprime} per channel. The combination of single- and double-tagged events is 
done by extending the likelihood function; the joint likelihood is the product of Poisson 
probabilities for each individual bin in each channel.
Systematic and statistical uncertainties  are incorporated  and 
treated as nuisance parameters with a Gaussian probability density function.
Figure~\ref{fig:limitsys} shows the observed and the expected limits from single- and double-tagged events 
combined.
Observed (expected) upper limits obtained on $\sigma(pp \rightarrow W'_R) \times {\cal {B}}(W'_R \rightarrow  tb)$ 
at 95$\%$ CL lie in the range  6.1 -- 1.0 (4.5 -- 1.4)~pb for $W'_R$ masses ranging from 0.5 to 2.0~\TeV. These  $\sigma \times {\cal {B}}$ 
limits are also applicable to a left-handed $W'$. 
The $\sigma \times {\cal {B}}$ limits are converted into mass limits using the intersection between the theoretical $\sigma \times {\cal {B}}$ curve 
as a function of $m_{W'_R}$ and the expected and observed $\sigma \times {\cal {B}}$ limit curves.
The corresponding observed (expected) 95$\%$ CL lower limit is: $m_{W'_R} > $ 1.13 (1.13)~\TeV. 
These are currently the most stringent direct limits on production of $W'_R \rightarrow tb$.

 \section{Acknowledgements}
 
 We thank CERN for the very successful operation of the LHC, as well as the
 support staff from our institutions without whom ATLAS could not be
 operated efficiently.
 
 We thank Z. Sullivan for discussions on the $W'$ model and for providing NLO signal cross section calculations.
 
 We acknowledge the support of ANPCyT, Argentina; YerPhI, Armenia; ARC,
 Australia; BMWF, Austria; ANAS, Azerbaijan; SSTC, Belarus; CNPq and FAPESP,
 Brazil; NSERC, NRC and CFI, Canada; CERN; CONICYT, Chile; CAS, MOST and NSFC,
 China; COLCIENCIAS, Colombia; MSMT CR, MPO CR and VSC CR, Czech Republic;
 DNRF, DNSRC and Lundbeck Foundation, Denmark; EPLANET and ERC, European Union;
 IN2P3-CNRS, CEA-DSM/IRFU, France; GNAS, Georgia; BMBF, DFG, HGF, MPG and AvH
 Foundation, Germany; GSRT, Greece; ISF, MINERVA, GIF, DIP and Benoziyo Center,
 Israel; INFN, Italy; MEXT and JSPS, Japan; CNRST, Morocco; FOM and NWO,
 Netherlands; RCN, Norway; MNiSW, Poland; GRICES and FCT, Portugal; MERYS
 (MECTS), Romania; MES of Russia and ROSATOM, Russian Federation; JINR; MSTD,
 Serbia; MSSR, Slovakia; ARRS and MVZT, Slovenia; DST/NRF, South Africa;
 MICINN, Spain; SRC and Wallenberg Foundation, Sweden; SER, SNSF and Cantons of
 Bern and Geneva, Switzerland; NSC, Taiwan; TAEK, Turkey; STFC, the Royal
 Society and Leverhulme Trust, United Kingdom; DOE and NSF, United States of
 America.
 
 The crucial computing support from all WLCG partners is acknowledged
 gratefully, in particular from CERN and the ATLAS Tier-1 facilities at
 TRIUMF (Canada), NDGF (Denmark, Norway, Sweden), CC-IN2P3 (France),
 KIT/GridKA (Germany), INFN-CNAF (Italy), NL-T1 (Netherlands), PIC (Spain),
 ASGC (Taiwan), RAL (UK) and BNL (USA) and in the Tier-2 facilities
 worldwide.

\onecolumngrid
\clearpage
% ATLAS Collaboration author list for 20-FEB-2012
% Data extracted on 15-Mar-2012 for paperid 204
%\documentclass[11pt]{article}
%\usepackage{a4wide}\begin{document}
\begin{flushleft}
{\Large The ATLAS Collaboration}

\bigskip

G.~Aad$^{\rm 48}$,
B.~Abbott$^{\rm 112}$,
J.~Abdallah$^{\rm 11}$,
S.~Abdel~Khalek$^{\rm 116}$,
A.A.~Abdelalim$^{\rm 49}$,
A.~Abdesselam$^{\rm 119}$,
O.~Abdinov$^{\rm 10}$,
B.~Abi$^{\rm 113}$,
M.~Abolins$^{\rm 89}$,
O.S.~AbouZeid$^{\rm 159}$
% ...
\bigskip

% ...
$^{10}$ Institute of Physics, Azerbaijan Academy of Sciences, Baku, Azerbaijan\\
$^{11}$ Institut de F\'isica d'Altes Energies and Departament de F\'isica de la Universitat Aut\`onoma  de Barcelona and ICREA, Barcelona, Spain\\
% ...
$^{48}$ Fakult\"{a}t f\"{u}r Mathematik und Physik, Albert-Ludwigs-Universit\"{a}t, Freiburg i.Br., Germany\\
$^{49}$ Section de Physique, Universit\'e de Gen\`eve, Geneva, Switzerland\\
% ...
$^{89}$ Department of Physics and Astronomy, Michigan State University, East Lansing MI, United States of America\\
% ...
$^{112}$ Homer L. Dodge Department of Physics and Astronomy, University of Oklahoma, Norman OK, United States of America\\
$^{113}$ Department of Physics, Oklahoma State University, Stillwater OK, United States of America\\
% ...
$^{116}$ LAL, Univ. Paris-Sud and CNRS/IN2P3, Orsay, France\\
% ...
$^{119}$ Department of Physics, Oxford University, Oxford, United Kingdom\\
% ...
$^{159}$ Department of Physics, University of Toronto, Toronto ON, Canada\\
% ...
\end{flushleft}

%\end{document}


\begin{thebibliography}{99}

\bibitem{ATLAS_det}
  ATLAS Collaboration, J. Inst. {\bf 3}, S08003 (2008).
  
\bibitem{Lumi}
  ATLAS Collaboration, Eur.\ Phys.\ J.\ C {\bf 71}, 1630 (2011)  [arXiv:1101.2185].

\bibitem{Lumi2}
  ATLAS Collaboration, ATLAS-CONF-2011-116, http://cdsweb.cern.ch/record/1376384.
  
\bibitem{extradimensions}
  A.~Datta, P.~J.~O'Donnell, Z.~H.~Lin, X.~Zhang and T.~Huang, Phys. Lett. B {\bf 483}, 203 (2000) [arXiv:hep-ph/0001059].
  
\bibitem{little-Higgs}
  M.~Perelstein, Prog. Part. Nucl. Phys. {\bf 58}, 247 (2007) [arXiv:hep-ph/0512128].
  
  
\bibitem{extended-gauge-model}
  G.~Altarelli, B.~Mele and M.~Ruiz-Altaba, Z. Phys. C {\bf 45}, 109 (1989).
  
\bibitem{Wprime_D0}
  D0 Collaboration, V.~M.~Abazov {\it et al.}, Phys. Lett. B {\bf 699}, 145 (2011) [arXiv:1101.0806]. 
  
\bibitem{Wprime_CDF}
  CDF Collaboration, T.~Aaltonen  {\it et al.}, Phys. Rev. Lett. {\bf 103}, 041801 (2009) [arXiv:0902.3276].
  
\bibitem{coordinates}
In the ATLAS coordinate system the pseudorapidity, $\eta$, is defined as $\eta = -\ln[\tan(\theta/2)]$, where $\theta$ is 
measured with respect to the $z$-axis, defined to be parallel to the beam. The azimuthal angle, $\phi$, is measured with respect to the $x$-axis, 
which points towards the center of the LHC ring, and the y-axis points upwards. 

  
\bibitem{ATLAS_trigger}
  ATLAS Collaboration, Eur.\ Phys.\ J.\ C  {\bf 72}, 1849 (2012) [arXiv:1110.1530].
  
  
\bibitem{pythia}
  T.~Sjostrand, S.~Mrenna and P.~Z.~Skands,
  JHEP {\bf 0605}, 026 (2006)
  [arXiv:hep-ph/0603175].
\bibitem{MRST2007}
  A.~Sherstnev and R.~S.~Thorne,
  Eur.\ Phys.\ J.\  C {\bf 55}, 553 (2008)
  [arXiv:0711.2473 [hep-ph]].


\bibitem{ZackSullivan}
  Z.~Sullivan, Phys.\ Rev.\ D {\bf 66}, 075011 (2002) [arXiv:hep-ph/0207290]

\bibitem{cteq6l}
  P.~M.~Nadolski {\it et al.}, Phys.\ Rev.\ D {\bf 78}, 013004 (2008) [arXiv:0802.0007]
  
  
\bibitem{SAMPLES-MCNLO1}
  S.~Frixione and B.~R.~Webber, JHEP {\bf 0206}, 029 (2002) [arXiv:hep-ph/0204244].

\bibitem{SAMPLES-MCNLO2}
  S.~Frixione, P.~Nason and B.~R.~Webber, JHEP {\bf 0308}, 007 (2003) [arXiv:hep-ph/0305252].
    
\bibitem{SAMPLES-HERWIG}
  G.~Corcella {\it et al.}, JHEP {\bf 0101}, 010 (2001) [arXiv:hep-ph/0011363].

\bibitem{JButterworth:1996zw}
  Butterworth {\it et~al.}, Z.\ Phys.\ C {\bf 72}, 637 (1996) [arXiv:hep-ph/9601371].


\bibitem{Aliev:2010}
  M.~Aliev {\it et al.}, Comput.\ Phys.\ Commun. {\bf 182}, 1034 (2011) [arXiv:1007.1327].

\bibitem{Martin:2009iq}
  A.~D.~Martin, W.~J.~Stirling, R.~S.~Thorne and G.~Watt, Eur.\ Phys.\ J.\ C {\bf 63}, 189 (2009) [arXiv:0901.0002].
  

 \bibitem{SAMPLES-ACER}
  B.~P.~Kersevan and E.~Richter-Was, arXiv:hep-ph/0405247 (2008).


\bibitem{Kidonakis:2010ux}
  N.~Kidonakis, Phys.\ Rev.\ D {\bf 82}, 054018 (2010) [arXiv:1005.4451].
  
\bibitem{Kidonakis:2010tc}
  N.~Kidonakis, Phys.\ Rev.\ D {\bf 81}, 054028 (2010) [arXiv:1001.5034].
\bibitem{Kidonakis:2011wy}
  N.~Kidonakis, Phys.\ Rev.\ D {\bf 83}, 091503 (2011) [arXiv:1103.2792].
  
  
   
  \bibitem{NLO-MCFM}
    J.~M.~Campbell, R.~K.~Ellis and C.~Williams, JHEP {\bf 1107}, 018 (2011) [arXiv:1105.0020].
    
  \bibitem{SAMPLES-ALPGEN} 
    M.~L.~Mangano, M.~Moretti, F.~Piccinini, R.~Pittau and A.~D.~Polosa, JHEP {\bf 0307}, 001 (2003) [arXiv:hep-ph/0206293].
    
  \bibitem{Anastasiou:2003ds}
    C.~Anastasiou, L.~J.~Dixon, K.~Melnikov and F.~Petriello, Phys.\ Rev.\ D {\bf 69}, 094008 (2004) [arXiv:hep-ph/0312266].
    
  \bibitem{Campbell}
    J.~Campbell and R.~K.~Ellis,  Phys.\ Rev.\ D {\bf 60}, 113006 (1999) [arXiv:hep-ph/9905386].

    
 \bibitem{ATL-2010-005}
 ATLAS Collaboration, Eur.\ Phys.\ J. {\bf C70}, 823 (2010) [arXiv:1005.4568].

 \bibitem{SAMPLES-G4}
  S.~Agostinelli, {\it et al.}, Nucl. Instrum. Meth. A {\bf 506}, 250 (2003). 
  
\bibitem{objects}
    ATLAS Collaboration, Eur.\ Phys.\ J.\ C {\bf 71}, 1577 (2011) [arXiv:1012.1792].
    
  \bibitem{muon_trig}
    ATLAS Collaboration, ATLAS-CONF-2011-021, https://cdsweb.cern.ch/record/1336750.
  \bibitem{muon_reso}
    ATLAS Collaboration, ATLAS-CONF-2011-046, https://cdsweb.cern.ch/record/1338575.
  \bibitem{muon_reco}
    ATLAS Collaboration, ATLAS-CONF-2011-063, https://cdsweb.cern.ch/record/1345743.

  \bibitem{ele_perf}
    ATLAS Collaboration, Eur.\ Phys.\ J.\ C {\bf 72}, 1909 (2012) [arXiv:1110.3174].


 
  \bibitem{Salam:2007xv}
    G.~P.~Salam and G.~Soyez, JHEP {\bf 0705}, 086 (2007) [arXiv:0704.0292]. 
    
  \bibitem{jetcali}
    ATLAS Collaboration, arXiv:1112.6426 (submitted  to Eur. Phys. J. C.)
    
    
    
  \bibitem{ATLAS-CONF-2011-102}
    ATLAS Collaboration, ATLAS-CONF-2011-102, https://cdsweb.cern.ch/record/1369219.
  
  \bibitem{ATLAS-CONF-2011-089}
    ATLAS Collaboration, ATLAS-CONF-2011-089, https://cdsweb.cern.ch/record/1356198.

  \bibitem{clusters}
    W.~Lampl {\it et al.}, ATL-LARG-PUB-2008-002 (2008), https://cdsweb.cern.ch/record/1099735.
    
  \bibitem{ATLAS-MET}
    ATLAS Collaboration, Eur.\ Phys.\ J.\ C  {\bf 72}, 1844 (2012) [arXiv:1108.5602].
    
  \bibitem{mtW}
    The $W$ boson transverse mass is defined as $m_{\mathrm{T}}(W)=\sqrt{2 p_{\rm T}^\ell E_\mathrm{T}^{\mathrm{miss}}(1 - \cos \Delta \phi) }$,
    where $p_{\rm T}^\ell$  is the $p_{\rm T}$ of the lepton and $\Delta \phi$ is the azimuthal angle separation between the lepton and $E_\mathrm{T}^{\mathrm{miss}}$.

  \bibitem{reduceQCD}    
    ATLAS Collaboration, Eur.\ Phys.\ J.\ C {\bf 71}, 1577 (2011) [arXiv:1012.1792].

  \bibitem{bumphunter}
    CDF Collaboration, T.~Aaltonen {\it et al.}, Phys.\ Rev.\ D {\bf 79}, 011101 (2009) [arXiv:0809.3781].
    
  \bibitem{SAMPLES-POWHEG}
    P.~Nason,
    JHEP {\bf 0411 } (2004)  040
    [arXiv:hep-ph/0409146].
    
  \bibitem{SAMPLES-POWHEG-1}
    S.~Frixione, P.~Nason, C.~Oleari,
    JHEP {\bf 0711 } (2007)  070
    [arXiv:hep-ph/0709.2092].

  \bibitem{Bayesian}
    A.~Caldwell, D.~Kollar and K.~Kr\"{o}ninger, Comput. Phys. Commun. {\bf 180}, 2197 (2009) [arXiv:0808.2552].
    
  \bibitem{Zwprime}
    ATLAS Collaboration, Phys.\ Lett.\ B {\bf 700}, 163 (2011) [arXiv:1103.6218].
    

\end{thebibliography}
\end{document}